\documentclass[reprint,amsmath,amssymb,aps,prl,twocolumn,showpacs,superscriptaddress]{revtex4-1}
\usepackage{amssymb}
\usepackage{graphicx}
\usepackage{dcolumn}
\usepackage{bm}
\usepackage{gensymb}

\begin{document}

\title{Measurement of Coherent Polarons in the Strongly Coupled Antiferromagnetically Ordered Iron-Chalcogenide Fe$_{1.02}$Te using Angle-Resolved Photoemission Spectroscopy}

\author{Z. K. Liu}
\affiliation{Stanford Institute for Materials and Energy Sciences,
SLAC National Accelerator Laboratory, Menlo Park, California 94025, USA}
\affiliation{Geballe Laboratory for Advanced Materials, Departments
of Physics and Applied Physics, Stanford University, Stanford,
California 94305, USA}
\author{R.-H. He}
\affiliation{Department of Physics, Boston College, Chestnut Hill, Massachusetts 02467, USA}
\author{D. H. Lu}
\affiliation{Stanford Synchrotron Radiation Lightsource, SLAC
National Accelerator Laboratory, Menlo Park, California 94025, USA}
\author{M. Yi}
\affiliation{Stanford Institute for Materials and Energy Sciences,
SLAC National Accelerator Laboratory, Menlo Park, California 94025, USA}
\affiliation{Geballe Laboratory for Advanced Materials, Departments
of Physics and Applied Physics, Stanford University, Stanford,
California 94305, USA}
\author{Y. L. Chen}
\affiliation{Department of Physics, University of Oxford, UK}
\affiliation{Stanford Institute for Materials and Energy Sciences,
SLAC National Accelerator Laboratory, Menlo Park, California 94025, USA}
\affiliation{Advanced Light Source, Lawrence Berkeley National
Laboratory, Berkeley, California 94720, USA}
\author{M. Hashimoto}
\affiliation{Stanford Synchrotron Radiation Lightsource, SLAC
National Accelerator Laboratory, Menlo Park, California 94025, USA}
\author{R. G. Moore}
\affiliation{Stanford Institute for Materials and Energy Sciences,
SLAC National Accelerator Laboratory, Menlo Park, California 94025, USA}
\author{S.-K. Mo}
\affiliation{Advanced Light Source, Lawrence Berkeley National
Laboratory, Berkeley, California 94720, USA}
\author{E. A. Nowadnick}
\affiliation{Stanford Institute for Materials and Energy Sciences,
SLAC National Accelerator Laboratory, Menlo Park, California 94025, USA}
\affiliation{Geballe Laboratory for Advanced Materials, Departments
of Physics and Applied Physics, Stanford University, Stanford,
California 94305, USA}
\author{J. Hu}
\affiliation{Department of Physics and Engineering Physics, Tulane
University, New Orleans, Louisiana 70118, USA}
\author{T. J. Liu}
\affiliation{Department of Physics and Engineering Physics, Tulane
University, New Orleans, Louisiana 70118, USA}
\author{Z. Q. Mao}
\affiliation{Department of Physics and Engineering Physics, Tulane
University, New Orleans, Louisiana 70118, USA}
\author{T. P. Devereaux}
\affiliation{Stanford Institute for Materials and Energy Sciences,
SLAC National Accelerator Laboratory, Menlo Park, California 94025, USA}
\author{Z. Hussain}
\affiliation{Advanced Light Source, Lawrence Berkeley National
Laboratory, Berkeley, California 94720, USA}
\author{Z.-X. Shen}
\affiliation{Stanford Institute for Materials and Energy Sciences,
SLAC National Accelerator Laboratory, Menlo Park, California 94025, USA}
\affiliation{Geballe Laboratory for Advanced Materials, Departments
of Physics and Applied Physics, Stanford University, Stanford,
California 94305, USA}

\date{\today}
\begin{abstract}
The nature of metallicity and the level of electronic correlations
in the antiferromagnetically ordered parent compounds are two
important open issues for the iron-based superconductivity. We
perform a temperature-dependent angle-resolved photoemission
spectroscopy study of Fe$_{1.02}$Te, the parent compound for iron
chalcogenide superconductors. Deep in the antiferromagnetic state,
the spectra exhibit a ``peak-dip-hump" line shape associated with
two clearly separate branches of dispersion, characteristics of
polarons seen in manganites and lightly-doped cuprates. As
temperature increases towards the Neel temperature (T$_N$), we
observe a decreasing renormalization of the peak dispersion and a
counterintuitive sharpening of the hump linewidth, suggestive of an
intimate connection between the weakening electron-phonon (e-ph)
coupling and antiferromagnetism. Our finding points to the
highly-correlated nature of Fe$_{1.02}$Te ground state featured by
strong interactions among the charge, spin and lattice and a good
metallicity plausibly contributed by the coherent polaron motion.

\end{abstract}

\pacs{74.25.Jb, 74.70.Xa, 79.60.-i, 71.38.-k}
\maketitle

The role of many-body interactions is one of the central questions
for unconventional superconductivity. For the recently discovered
iron-based superconductors, the strength of electronic correlations
is still an unsettled issue \cite{pnictidecorrelation,dhreview}. For
one of them, iron chalcogenides, a strong correlation scenario has
been proposed by theory
\cite{theoryvishwanath,theorylowerhubbardband} and supported by
experiments
\cite{felixft42arpes,takahashift30arpes,donglaiftarpes,guopticalcorrelation,ikedafesecorrelation,specificheatmao,phasediagrammao,opticalwang}.
For their parent compound Fe$_{1+y}$Te, while the high-temperature
paramagnetic (PM) state shows similar signs for localized physics as
in the undoped high-Tc cuprates in transport \cite{phasediagrammao}
and optical \cite{opticalwang} experiments, the metallic behavior in
the low-temperature antiferromagnetic (AFM) state (at T$<$T$_N$,
T$_N$=72 K for y=0.02) \cite{phasediagrammao,samplegrowthmao} seems,
\emph{prima facie}, to deviate from localized physics and questions
the importance of strong correlations.

\begin{figure*}[!]
\includegraphics[width=1.5\columnwidth]{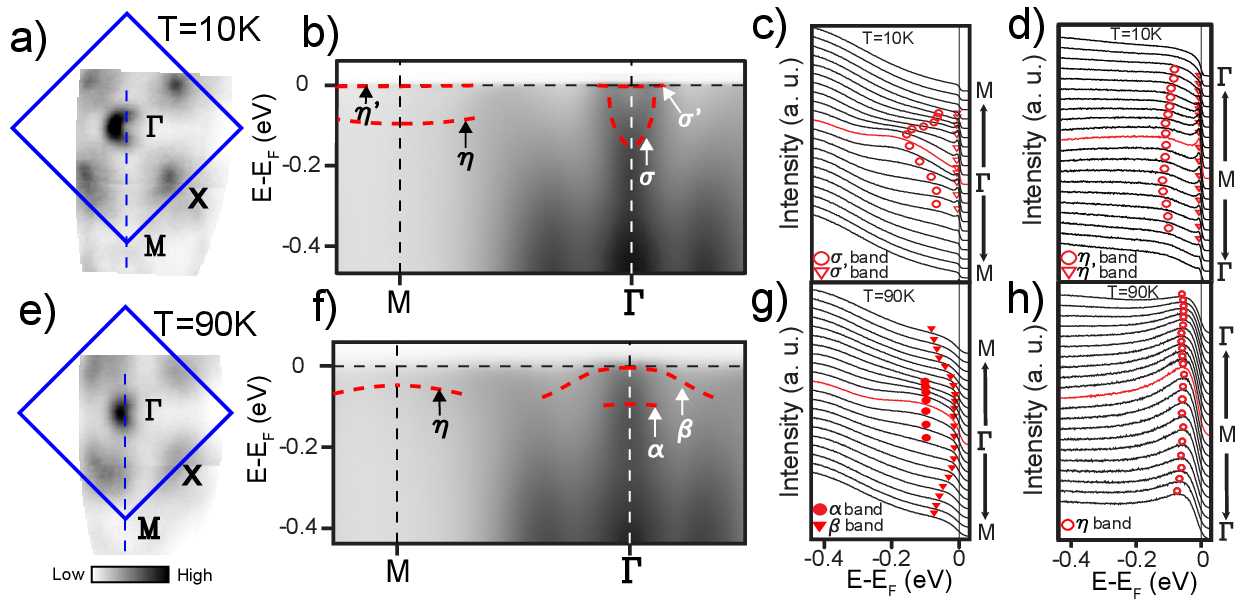}
\caption{(a),(e) Fermi surface map of Fe$_{1.02}$Te measured with 22
eV excitation energy at (a)T=10 K and (e)T=90 K. The photoemission
intensity is integrated over 10 meV window around the E$_F$. (b),(f)
Photoemission intensity of the cut along the $\Gamma$-M direction of
(a) and (e), respectively. Dashed curves are eyeguides of
dispersion. (c),(g) Plot of the energy distribution curves (EDCs)
around $\Gamma$ along the cuts in (b) and (f). Labeling marks are
local maxima of the EDCs after dividing the corresponding
Fermi-Dirac function and background subtraction (see below). (d),(h)
Plot of the EDCs around M along the cuts in (b) and (f). Labeling
marks are local maxima of the EDCs.}
\end{figure*}

In terms of the strength of coupling between itinerant electrons and
other degrees of freedom including the localized spins, a recent
theoretical work \cite{kumagneticcorrelation} has pointed out
similarities between iron chalcogenides and colossal
magnetoresistive (CMR) manganites, a strongly-correlated prototype
system also with a (ferro)magnetically ordered metallic ground
state. A salient manifestation of strong coupling with collective
modes in CMR manganites is the self-energy effect seen in the
single-particle spectral function measured by angle-resolved
photoemission spectroscopy (ARPES), which is characterized by a
characteristic ``peak-dip-hump" line shape that has been attributed
to polaron formation
\cite{normanmanganitearpes,normanpolaroncondensation}. In this
letter, we present temperature-dependent ARPES study on
Fe$_{1.02}$Te. Our result shows that the spectra in the AFM state
contain signature of polarons reminiscent of those found in CMR
manganites \cite{normanmanganitearpes, dessaumanganitearpes} and
deeply underdoped cuprates \cite{kylepolaron1, shenarpesrev}. This
observation thus raises an intriguing perspective that the good
metallicity of Fe$_{1.02}$Te at low temperature arises from coherent
polaron motion, as proposed for the manganites case
\cite{normanpolaroncondensation}. Different from the manganites
case, however, the temperature evolution of the polaron feature
shows signs of concomitant weakening of the strong-coupling polaron
behavior and the magnetic ordering upon the increase of temperature.
This in turn suggests the electronic correlations likely strengthen,
rather than weaken as generally thought, in the AFM state. The
observed intimate tracking of polaron behavior with the magnetic
ordering points to a cooperation between lattice and magnetism as a
key factor driving the low-temperature system towards the
strong-coupling limit.

High quality Fe$_{1.02}$Te single crystals were synthesized using
flux method \cite{samplegrowthmao}. Excess Fe ratio was kept as low
as possible and was determined by energy-dispersive X-ray
spectrometry to be around 2$\%$. ARPES measurements were performed
at beamline 5-4 at Stanford Synchrotron Radiation Lightsource
(photon energy $h\nu=22$ eV). The energy(angle) resolution is 7
meV(0.3$^\circ$). The samples were cleaved \emph{in situ}, and
measured in ultrahigh vacuum with pressure better than
$3.0\times10^{-11}$ Torr.

We first compare the electronic structure of Fe$_{1.02}$Te above (90
K) and below (10 K) the AFM transition (Fig. 1). The electronic
structure in the PM state [Fig. 1(e)-(h)] is characterized by
overall broad features. Along the high symmetry $\Gamma$-M direction
in 2-Fe unit cell Brillouin zone, we can identify two hole-like
bands ($\alpha$, $\beta$) around $\Gamma$ and one
hole-like band ($\eta$) around M. %The dispersion are denoted by dashed curves on the photoemission intensity plots in Fig. 1(b2) and labels for peaks on the EDCs in Fig. 1(c2)-(d2).
The observed band dispersions show partial agreement with the DFT
calculation \cite{dftdavid}: the $\beta$ and $\eta$ bands roughly
follow the calculated dispersion, with the calculated bandwidth
renormalized by a factor of 5. The predicted outermost hole-like
band at $\Gamma$ and electron-like band at M may be suppressed by
the polarization matrix elements. The photoemission intensity
observed around X as shown on the Fermi surface plot [Fig. 1(e)] is
not predicted by the calculation. We note that we do not see
well-defined hole-like band duplicating $\Gamma$ feature at X as
previously reported in Ref. \cite{hasanftarpes}. Our spectra would
be similar to those in Ref. \cite{donglaiftarpes} if their Brillouin
zone definition is rotated by 45$^o$ \cite{[{}][{We used X-ray
diffraction to align the samples, in such a way that is justified by
a comprehensive Se-concentration dependence study of the band
structure of Fe$_{1+y}$Se$_x$Te$_{1-x}$, which will be reported
elsewhere.}]liunextpaper}.

Comparing with the PM state, electronic structure of Fe$_{1.02}$Te
in the AFM state is drastically different [Fig. 1(a)-(d)]: One
electron-like feature is identified around the $\Gamma$ point
($\sigma$ band). The $\eta$ band at M shifts further away from
E$_F$. The $\sigma$ and $\eta$ bands are characterized by very broad
humps in EDCs and do not appear to cross E$_F$. In the vicinity of
E$_F$, sharp quasiparticle peaks with small spectral weight are
observed at both $\Gamma$ ($\sigma$' band) and M ($\eta$' band).
Note that these two sharp quasiparticle bands are not predicted in
the bandstructure calculation for the AFM state \cite{xiangtaodft},
nor do they look like extrinsic effects (such as impurity induced
features) since they only appear close to E$_F$ where $\sigma$ and
$\eta$ features are observed.

We next focus on the $\sigma$ and $\sigma$' bands inspired by their
intimate dispersion relationship observed [Fig. 2(a)-(c)]. To track
the features close to and above E$_F$, we divide each EDC by the
corresponding Fermi-Dirac function at the measurement temperature
convolved with the instrument resolution [Fig. 2(a)]. We then
perform background subtraction to highlight the $\sigma$ and
$\sigma$' features. An EDC far away from $\Gamma$ where $\sigma$ and
$\sigma$' bands both have vanishing intensity is chosen as the
background and subtracted from all the EDCs around $\Gamma$
\cite{notebg}.

The EDC plot of the $\sigma$ and $\sigma$' bands [Fig. 2(b)] show
canonical two-pole spectral functions, commonly referred to as the
peak-dip-hump line shape \cite{shenarpesrev}. Local minima (the
dips) are observed at 18 meV below E$_F$ and break the dispersion
into two branches. The high energy branch, the $\sigma$ band, shows
a broad hump feature which can be well fitted by a Gaussian
function. The maxima of the hump overall follow the band dispersion
determined by a parabolic fitting of the momentum distribution curve
(MDC) peaks [Fig. 2(c)]. But it starts to deviate from the MDC
derived dispersion, levels off and tends to bend back when getting
close to around 60 meV below E$_F$. The low energy branch,
electron-like $\sigma$' band, is characterized by a sharp
quasiparticle peak and could be well fitted by a Lorentzian
function. It also has small bandwidth: A parabolic fitting shows its
effective mass of $\sim$18 m$_e$ at 30 K, which is $\sim$90 times
larger than the band mass derived from the MDC dispersion, which was
previously demonstrated to produce a band dispersion akin to the LDA
predicted bareband \cite{normanmanganitearpes}. Similar features are
also observed in the $\eta$ and $\eta$' bands at the M point (see
below).

\begin{figure}[!]
\includegraphics[width=\columnwidth]{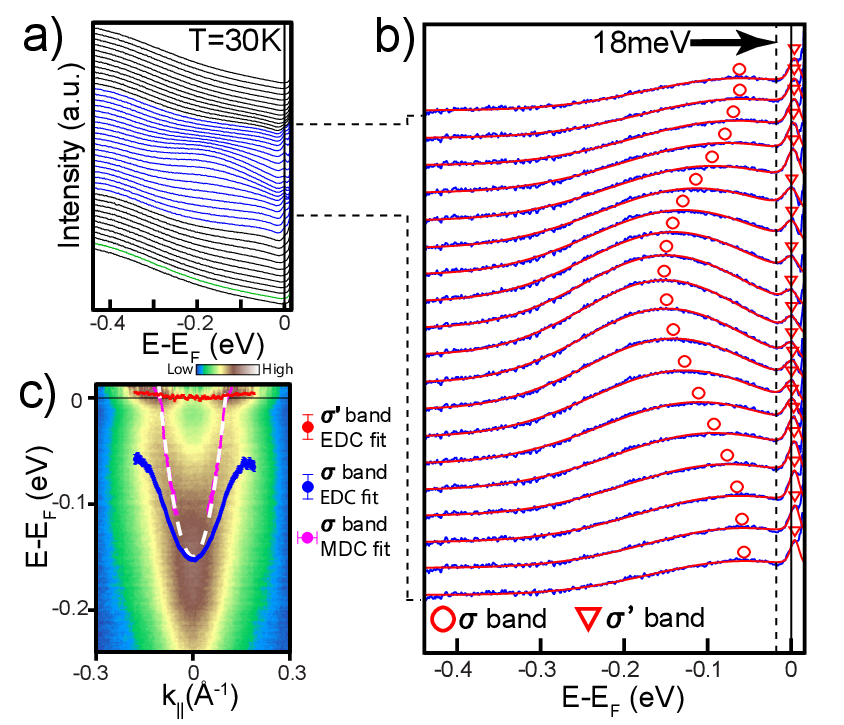}
\caption{(a) Plot of the EDCs around $\Gamma$ along the $\Gamma$-M
direction cut measured at 30 K. The data is plotted after
normalizing to the intensity at the highest binding energy and
dividing the corresponding Fermi-Dirac function. The green EDC is
taken as the background to be subtracted from the blue EDCs around
$\Gamma$. (b) Plot of the blue EDCs around $\Gamma$ after background
subtraction in (a). Red lines are Gaussian+Lorentzian fitting
results of the EDCs. Fitted Guassian(Lorentzian) peaks are marked by
open circles(triangles). The dashed line denotes the position of the
dips in the EDCs, which is 18 meV below E$_{F}$. (c) Photoemission
intensity plot of the EDCs in (b), together with marks labeling the
peak positions in the EDCs and MDCs. Peak positions of the MDCs of
the $\sigma$ band are determined by fitting to a two-Lorentzian
function. The dashed curve is the parabolic fit of the $\sigma$ band
MDC peaks.}
\end{figure}

Such self-energy effect in the single-particle spectral function of
Fe$_{1.02}$Te bears strong resemblance to that seen in deeply
underdoped cuprates \cite{kylepolaron1, shenarpesrev} and CMR
manganites La$_{2-2x}$Sr$_{1+2x}$Mn$_2$O$_7$
\cite{normanmanganitearpes, dessaumanganitearpes}. A widely-accepted
interpretation for those features in cuprates and manganites is due
to the strong coupling between electrons and some bosonic collective
modes, which leads to the formation of, \emph{e.g.}, polarons in the
case of manganites. In this scenario, the hump feature describes the
incoherent excitations of electrons strongly coupled to a bath of
bosons (phonons) and the small quasiparticle peak which forms a
heavily renormalized band associated with the coherent polaron
motion \cite{normanpolaroncondensation}. Our observation of the
peak-dip-hump structure in the spectra and large effective mass
enhancement of the quasiparticle band in Fe$_{1.02}$Te is consistent
with the polaron interpretation. In such a picture, the energy scale
of the involved collective mode can be estimated from the dip
position in the EDCs to be about 18 meV, which is very close in
energy to the A$_{1g}$ phonon mode observed in Raman spectroscopy
\cite{zhangphonon,takagiphonon} but rather different from the
reported ($\pi$,0) magnetic resonance mode at $\sim$7 meV
\cite{magnonenergy,magnonenergy2}. This comparison suggests that the
phonon is more likely the direct agent involved in the polaron
formation, but as we will see below that the e-ph coupling alone
might not be sufficient.

\begin{figure}[!]
\includegraphics[width=\columnwidth]{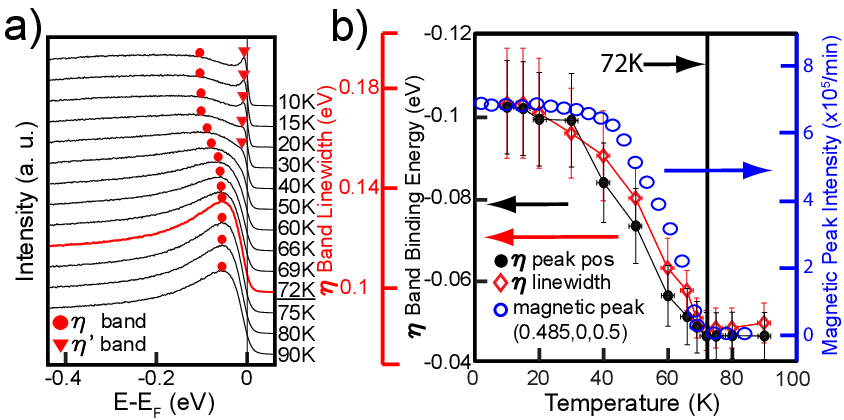}
\caption{(a) Plot of the EDCs at the M point at various
temperatures. The red EDC is recorded when T=T$_N$. Marks labeling
peaks of the $\eta$ and $\eta$' bands are local maxima of the EDCs.
(b) Plot of the $\eta$ band binding energy and linewidth at the M
point together with the (0.5, 0, 0.5) AFM Bragg peak intensity
versus temperature. The magnetic peak intensity curve is adapted
from a published neutron scattering experiment \cite{magcurve} and
roughly proportional to the ordered magnetic moment of Fe.}
\end{figure}

The temperature evolution of the ARPES spectra, especially across
the AFM to PM phase transition, provides deeper insights into the
polaron scenario in Fe$_{1.02}$Te. In Fig. 3(a), we show the M point
EDCs at various temperatures. As temperature increases, we observe
distinct evolution behavior of the hump ($\eta$ band) and the peak
($\eta$' band) features: the quasiparticle peak in the $\eta$' band
quickly loses spectral weight and becomes indiscernible eventually
for T$>$50 K; Meanwhile, the peak in the $\eta$ band first stays
almost unchanged below 30 K. At 30 K$<$T$<$T$_N$, the maximum
position shifts towards lower binding energy (BE) and the linewidth
of the hump becomes narrower. Finally above T$_N$, the $\eta$ band
stays basically unchanged again.

The distinct behavior of the $\eta$ and $\eta$' bands together
reveals how polarons evolve with temperature. A similar spectral
weight reduction of the quasiparticle peak is also observed in the
temperature evolution of the polaron line shape in manganites
\cite{normanpolaroncondensation, lucpolarontemperature} and was
interpreted therein as loss of coherence of condensed polarons. The
motion of coherent polarons at low temperature has been proposed to
be an important factor (in addition to the double exchange
mechanism) that contributes to the low-temperature metallicity of
manganites. The observed temperature dependence of the $\eta$' band
is consistent with the polaron scenario and, by analogy, we propose
that the coherent polaron motion might also play an important role
in the metallic transport in the AFM state of Fe$_{1.02}$Te - an
important possibility that, to our knowledge, has been overlooked so
far.

\begin{figure}[!]
\includegraphics[width=\columnwidth]{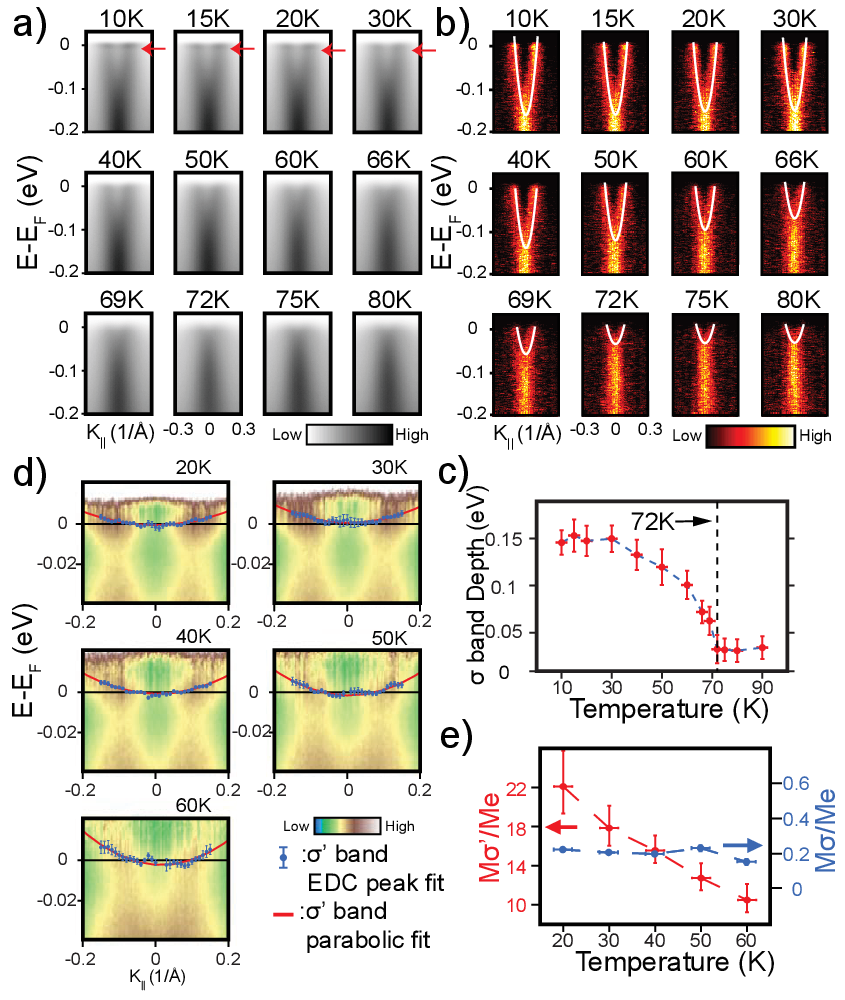}
\caption{(a) Photoemission intensity around $\Gamma$ along the $\Gamma$-M direction at various temperatures. Red arrows indicate the positions of the dips in the spectra. (b) The second derivative plot of each intensity plot in (a). White curves are the parabolic fit of the MDC peaks found in (a). (c) Plot of the $\sigma$ band depth (derived from the fitting results in (b)) versus temperature. (d) Magnified plot of photoemission intensity in the E$_F$ vicinity at various temperatures. The data are processed using the similar method as in Fig. 2. Blue marks denote the peaks of the EDCs of the $\sigma$' band and the red curve is the parabolic fit of these peaks. (e) Plot of the effective mass of the $\sigma$' and $\sigma$ band versus temperature. The effective mass of the $\sigma$' and $\sigma$ band is calculated from the EDC and MDC [Fig. 4(b)] fitting results, respectively. If not shown, the error bars are smaller than the symbol's size.}
\end{figure}

The evolution of the hump ($\eta$ band) feature shows some onset
behaviors at the magnetic ordering transition, different from the
manganite case. In manganites the humps are broader and shift to
higher BE at higher temperatures \cite{normanpolaroncondensation,
lucpolarontemperature}. In contrast, in Fe$_{1.02}$Te, the humps get
narrower at higher temperatures [Fig. 3(a)] - a trend opposite to
the expectation for the mere thermal smearing \cite{makototb} - and
shift toward low BE as temperature increases. We plot the BE and the
hump linewidth of the $\eta$ band together with the Fe magnetic
moment as a function of temperature in Fig. 3(b) and find that all
of them show concomitant changes tied to T$_N$. The linewidth change
of the hump shows the weakening of electron incoherence (likely by
phonon scattering) as Fe magnetism decreases rapidly across T$_N$.
While the observed band shift is certainly related to the AFM
ordering, it cannot be directly explained by the resulting band
reconstruction, because the ordering vector is in $\Gamma$-X
direction instead of $\Gamma$-M and the bandstructure calculation
did not reproduce the observed shift \cite{xiangtaodft}.
Alternatively, this apparent band shift could be taken as a natural
consequence of the disappearing of the EDC ``dip" that sets the peak
and hump apart at low temperatures but can no longer be clearly
resolved at T$>$50 K. Therefore, the entire evolution of the $\eta$
band hump likely suggests the dissociation, rather than decoherence,
of polarons, as a result of a weakened e-ph coupling upon
approaching the magnetic ordering transition, which does not seem to
occur in manganites.

Such a unique aspect of the polaron formation in Fe$_{1.02}$Te is
further supported by a similar temperature evolution of the $\sigma$
and $\sigma$' bands observed at the $\Gamma$ point, despite the
complications therein introduced by the band reconstruction due to
the AFM ordering (Fig. 4; see the supplemental material for a detailed discussion): the $\sigma$ band shifts up from $\sim$150meV to $\sim$30meV below E$_F$ and becomes a part of the $\beta$ band around $\Gamma$ as the temperature increases, while the ``vertical dispersion" sitting at the $\Gamma$ point [Fig. 4(a)$\&$(b)] becomes more prominent at high temperatures and is identified to be the $\alpha$ band. The existence of the $\sigma$' band is indicated by the red arrows in Fig. 4(a) pointing to positions where the ARPES spectra break up into two dominant parts (the dips). Up to 30K the dips are clearly discernible and the positions unchanged, whereas they
become increasingly obscure upon raising temperature. At T$>$60 K
(at $\Gamma$), both branches merge into one. Additionally, we could
extract the effective mass of the $\sigma$' band from detailed EDC
analysis at temperatures where the $\sigma$' band is discernable
[Fig. 4(d)]. While the effective mass of the $\sigma$ band does not
show significant variation, the effective mass of the $\sigma$' band
decreases as the temperature increases [Fig. 4(e)]. Such observation
at $\Gamma$ provides a complementary angle to see how the e-ph
coupling decreases when the AFM order diminishes.

Taken collectively, the observed temperature evolutions of the
polaron features at both $\Gamma$ and M suggest that the e-ph
coupling weakens along with the demise of the AFM. Consistent with
these, a recent Raman experiment shows that the linewidth of the
characteristic A$_{1g}$ phonon mode of the appropriate energy of the
dip ($\sim$20 meV) is broader at low temperature and narrower at
high temperature, and the change is most dramatic across T$_N$
\cite{modesoft}.

From a theoretical perspective, antiferromagnetism could either reduce or increase the critical e-ph interaction for a polaron crossover. On one hand, carriers are slowed down due to surrounding spin flip clouds which make them subject to stronger e-ph interactions and a polaron formation at a smaller critical coupling; however, strong electronic correlations needed for antiferromagnetism can suppress charge fluctuations and the associated e-ph interaction, which would make polaron formation more difficult. This problem has been studied with several approaches in the context of the underdoped cuprates. Diagrammatic quantum Monte Carlo studies of a single hole in the t-J model coupled to optical phonons found that antiferromagnetism reduced the critical e-ph coupling for polaron formation \cite{nagaosa_pol}. In contrast, dynamical mean field theory studies of polaron formation in the Hubbard-Holstein model have found an increase in the critical e-ph coupling for polaron formation in both PM \cite{pm_dmft} and AFM \cite{afm_dmft} state, yet the increase is much smaller in AFM state. A study utilizing the dynamic cluster approximation has found a synergistic interplay between antiferromagnetism and polaron formation, and a reduction in the critical coupling for polaron formation \cite{dym_cluster}. These theoretical proposals suggest that the presence of antiferromagnetism helps polaron formation, compatible with our observations. Such a picture of the polaron formation as the result of a cooperative interplay among the magnetism and e-ph coupling sets Fe$_{1.02}$Te uniquely apart from manganites.

We thank A. S. Mishchenko, A. F. Kemper, B. Moritz, D. J. Singh and
J. S. Wen for enlightening discussions. This work is supported by
the Department of Energy, Office of Basic Energy Sciences, Division
of Materials Science. The work at Tulane is supported by the NSF
under Grant No. DMR-0645305 and the LA-SiGMA program under Award No.
EPS-1003897.

\bibliography{liu_fete_arpes_main}
\end{document}